\documentclass[manuscript]{aastex}
\begin{document}

\title{Rotational quenching of CO due to H$_2$ collisions}
\author{Benhui Yang and P. C. Stancil}
\affil{Department of Physics and Astronomy and the Center for
     Simulational Physics,\\  The University of Georgia,
         Athens, GA 30602-2451}
\email{yang@physast.uga.edu, stancil@physast.uga.edu}
\author{N. Balakrishnan}
\affil{Department of Chemistry, The University of Nevada Las Vegas, 
 Las Vegas, NV 89154}
\email{naduvala@unlv.nevada.edu}
\author{R. C. Forrey}
\affil{Department of Physics, Penn State University,
Berks Campus, Reading, PA 19610} 
\email{rcf6@psu.edu}

\begin{abstract}
Rate coefficients for state-to-state rotational transitions in CO induced by 
both para- and ortho-H$_2$ collisions are presented. The results were obtained
using the close-coupling method and the coupled-states approximation,
with the CO-H$_2$ interaction potential of Jankowski \& Szalewicz (2005).
Rate coefficients are presented for temperatures between 1 and 3000 K,
and for CO($v=0,j$)  quenching from $j=1-40$ to all lower $j^\prime$ levels.
Comparisons with previous calculations using an earlier potential
show some discrepancies, especially at low temperatures and for 
rotational transitions involving large $|\Delta j|$.
The differences in the well depths of 
the van der Waals interactions and the anisotropy of the two potential surfaces
lead to different resonance structures in the energy dependence
of the cross sections which influence the low temperature
rate coefficients. Applications to far infrared observations of astrophysical
environments are briefly discussed. 
\end{abstract}


\section {INTRODUCTION}

Molecular hydrogen and carbon monoxide are the most abundant molecular species 
in the majority of interstellar environments,
and they play important roles in determining  the physics and chemistry of 
diffuse clouds \citep{herb01}.
CO is observed in diffuse clouds in absorption in the ultraviolet (UV) and visible and 
in emission at
 infrared (IR) wavelengths
\citep{snow06}, revealing the  physical and chemical complexity 
along these lines of sight. As an example, \citet{liszt06,liszt07} recently
investigated the 
formation, fractionation, 
and rotational excitation of CO in H$_2$ bearing diffuse H I clouds.
Due to its small rotational constant, CO can be easily 
rotationally excited by collisions with other species in interstellar 
gas, mostly H$_2$, with CO rotational lines providing important  
diagnostics of gas density and temperature.  
The abundance ratio, CO/H$_2$, is assumed to be roughly 
constant in dense molecular gas, as such, 
CO is often used as a tracer of H$_2$, as the latter is difficult to
detect in emission in cold environments. 
The observed CO abundance is therefore  used to estimate 
the total H$_2$ content \citep{sonn07}.

In environments with an intense UV field, the radiation can drive the
chemistry and internal level populations out of equilibrium. In such
situations, a
photodissociation region (PDR) resides at the interface of the hot H II
region and the cold molecular region.  
In recent years various codes 
 which model the physics and chemistry of  PDRs have
been developed with an emphasis on the excitation mechanisms of
both H$_2$ and CO \citep{dish88,warin96,lep06,sha05,rollig07}.

In more dense environments such as low-mass dwarf stars (i.e., M, L, or T dwarfs),  CO is an important opacity source and
its infrared line list can be obtained quite accurately \citep{pavl02,jones05}. It has also
been observed in the dayside spectrum of the transiting hot-Jupiter extrasolar giant
planet (EGP) HD 189733b \citep{swa09}. While molecular level populations are typically treated
in local thermodynamic equilibrium (LTE) in stellar and planetary atmospheres,
\citet{sch00} have pointed out that non-LTE conditions may exist in very cool stellar
atmospheres or in cases where the temperature of the incident radiation, i.e. close-in EGPs,
is much different from the kinetic temperature of the gas. 

Due to the astrophysical importance of molecular hydrogen and carbon
monoxide, the CO-H$_2$ collisional system has 
been the subject of numerous experimental 
\citep{kudian91,butz71,kupper73,boua73,nerf75,brech80,andre82,picar83,
schra91,dras98,mcke90,mcke91,mcke98, anto00} 
and  theoretical \citep{green76, pris78, flower79,  
poul82,schin84, bacic85, bacic852, jankow98, jankow05, hemert83,  
parish92, danby93,sala95, reid97,meng01,flower84,flower85,flower01, 
anto00, wern06, yang061, yang062} studies. 
Quantitative determinations of state-to-state cross sections and rate
coefficients for CO-H$_2$ collisions are crucial to numerical 
models of various astrophysical environments, such as those highlighted
above. 
However, as measurements of these quantities are difficult, 
numerical models often rely on cross sections 
and rate coefficients derived from theoretical methods. In a highly-cited work, 
\citet{green76}  performed close-coupling 
calculations of rate 
coefficients based on an approximate CO-H$_2$ potential surface. Since then,
a number of quantum scattering calculations were
carried out on various potential energy surfaces (PESs). In this work,
we have performed comprehensive calculations of state-to-state
cross sections in an effort to develop a complete database of rotational
quenching rate coefficients, but utilizing a more recent, and presumably
more accurate PES. In the
following sections, the choice of the potential surface, the adopted
scattering approach, and the results are discussed. We briefly highlight 
important astrophysical applications of the current CO-H$_2$ rate
coefficients for modeling of far IR (FIR) observations. 

\section{THE POTENTIAL ENERGY SURFACE}

 A number of PESs \citep{pris78, flower79, poul82, schin84, bacic85, jankow98, jankow05} 
have been developed for the CO-H$_2$ complex. 
One of the most accurate surfaces was given by \citet{jankow98}
who calculated a four-dimensional, rigid-rotor 
PES referred to as V$_{98}$.  
This surface was constructed using the symmetry-adapted perturbation theory 
(SAPT) method  with high-level electron correlation effects. 
An analytical fit of the {\it ab initio}  PES
has a global minimum of 109.3 cm$^{-1}$ at the intermolecular 
separation of 7.76~a$_0$  for the linear geometry with the C atom pointing 
toward H$_2$.
\citet{anto00} adopted V$_{98}$ in their calculations of  state-to-state 
cross sections for CO rotational
excitation at collision energies  between 795 and 991 cm$^{-1}$  which
were found to be in good agreement with their measurements. 
The V$_{98}$ PES was also utilized in 
second virial coefficient calculations for mixtures of hydrogen and CO \citep{gott00}. 
A comparison with experimental data suggested that
the van der Waals well of the V$_{98}$ PES 
is too deep by 
4\%-9\%, though it represents an improvement over earlier surfaces.
The V$_{98}$ surface has also been used in full coupled-channel cross section 
and rate coefficient calculations for rotationally inelastic scattering of 
CO by  para- and ortho-H$_2$ \citep{flower01,meng01}.  
However, because the attractive well of V$_{98}$  
was presumed to be too  deep, \citet{meng01} modified the PES by  
merely multiplying the surface by a constant factor of 0.93 which was
subsequently used in their scattering calculations. 

A newer CO-H$_2$ PES was reported by \citet{jankow05}, 
referred to as V$_{04}$. To improve the surface accuracy, they adopted the 
coupled-cluster method with single, double, and 
noniterative triple excitations [CCSD(T)] and the supermolecular approach.
The V$_{04}$ surface was 
obtained on a five-dimensional grid including the 
dependence on the H-H separation, but with the CO molecule taken to be rigid 
with the C-O separation set to the value of the C-O distance 
averaged over the CO ground state vibrational wave function. The PES was 
then obtained by averaging over the intramolecular vibration of H$_2$ to
yield a four-dimensional rigid-rotor surface. 
Similar to V$_{98}$, an analytical fit of the V$_{04}$ was presented 
which has a global minimum of 93.05 cm$^{-1}$, or 16.22 cm$^{-1}$ shallower
than $V_{98}$.
\citet{jankow05} used V$_{04}$ to calculate the rovibrational energy levels
for the para- and ortho-H$_2$-CO complex as well as the second virial coefficient.
The rovibrational energies were found to agree with the experimental values of
\citet{mcke98} for para-H$_2$ to within 0.1 cm$^{-1}$ while a scaling of V$_{04}$,
resulting in a deepening of the well depth by 4 cm$^{-1}$, was needed to match
the measured second virial coefficients.
V$_{04}$ may be too shallow though it is possible that the experimental
second virial coefficients are systematically too low \citep{jankow05}.
Comparative studies of collisional cross sections using the two
surfaces have been presented in our earlier work \citep{yang061} and by
\citet{wern06}. In \citet{yang061}, calculations with the V$_{04}$ surface
gave cross sections in excellent agreement with the state-to-state
measurements of \citet{anto00}.

\section{THEORETICAL APPROACH}

The theory for the scattering of two linear rigid rotors has been developed  
in \citet{green75} and \citet{heil78}.
The calculations presented here were performed for collision
energies between 10$^{-5}$ and 10,000 cm$^{-1}$ by applying 
both the close-coupling (CC) method
and the coupled-states (CS) approximation. All the CC  and CS calculations were performed using 
the nonreactive scattering code MOLSCAT \citep{molscat94}. 
In the quantum scattering calculations, 
the coupled-channel equations were integrated using the modified
log-derivative Airy propagator of \citet{alex86}
with a variable step size. 
The largest Legendre terms adopted in the potential expansion for H$_2$ and CO
were, respectively,  8 and 10. The numbers of Gauss integration points
used in projecting angular components of the potential were 10, 11, and 12
for integration in $\theta_1$, $\theta_2$, and $\phi$, respectively.
The propagation was carried out to a maximum intermolecular separation of
$R=100$  \AA.  At each energy, a sufficient
number of total angular momentum partial waves was included to ensure
convergence of the state-to-state cross sections to within 1\%. 
The maximum value of the total angular
momentum quantum number $J$ employed in the
calculations was 300.
Since, the computation time for CC calculations scales as $j_{\rm max}^6$, the CS
method is adopted for energies greater than 800 cm$^{-1}$ which results in
considerable  time savings with a scaling proportional to  $j_{\rm max}^4$, 
where $j_{\rm max}$ is the maximum size of the rotational basis. Further
details about the scattering calculations can be found in \citet{yang061,yang062}.
The adopted rotational constants for H$_2$ and CO 
are 60.853 cm$^{-1}$ \citep{hub79} and 1.9225 cm$^{-1}$ \citep{lov79}, respectively.
 Hereafter, $j_1$ denotes the rotational quantum
number for H$_2$, and likewise $j_2$ for CO. 
In the present study, we adopted the four-dimensional PES for the
CO-H$_2$ complex,  V$_{04}$,  of \citet{jankow05}.

Calculations of the collision energy dependence of
state-to-state quenching cross sections for initial rotational states of
CO, $j_2$=1, 2, $\cdots$, 40,  by collisions with both para- and ortho-H$_2$
(i.e., $j_1=0,1$, respectively)
were performed. The cross sections  were
thermally averaged over the kinetic energy distribution to yield state-to-state
rate coefficients  from specific initial rotational states $j_2$  
as functions of the temperature $T$, 

\begin{equation}
k_{j_2\rightarrow j_2'}(T)= \left (\frac{8}{\pi\mu\beta} \right )^{1/2}\beta^2\int^{\infty}_0 E_k
\sigma_{j_2\rightarrow j_2'}(E_k)\exp(-\beta E_k)dE_k
\end{equation}
where $\sigma_{j_2\rightarrow j_2'}$ is the rotational transition cross section with
$j_2$ and $j_2'$ being respectively the initial and final rotational quantum  number of
CO, $\mu$ is the reduced mass of the CO-H$_2$ complex, $E_k$ the kinetic energy, and
$\beta=(k_BT)^{-1}$, where $k_B$ is the Boltzmann constant.

\section{RESULTS AND DISCUSSION}

\begin{figure}
\epsscale{1.0}
\plotone{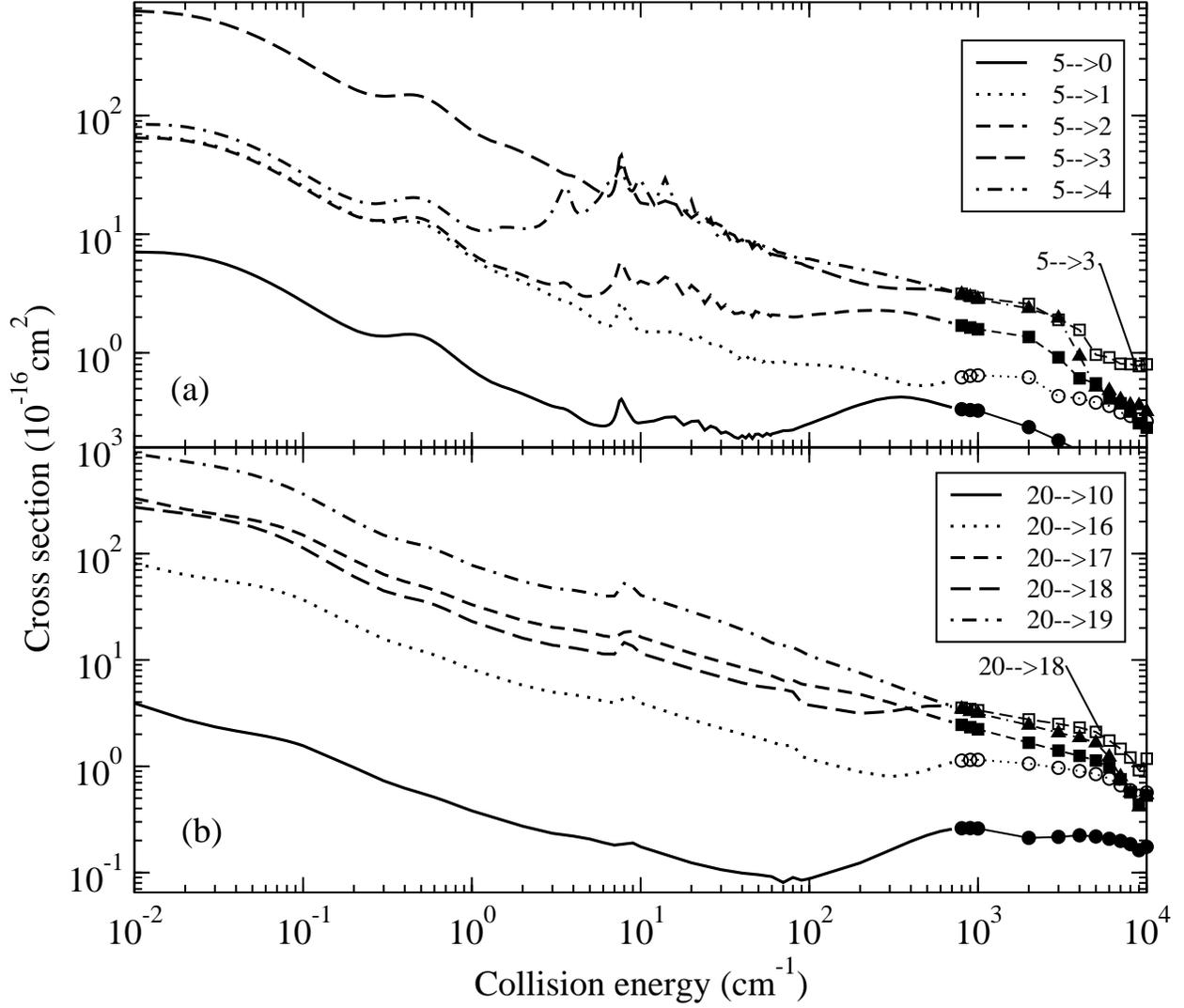}
\caption{Cross sections for the quenching of CO($j_2$)
due to para-H$_2$ collisions as a function of collision energy
obtained with the 
close-coupling method (lines) and coupled-states approximation (symbols).
(a) $j_2=5$, (b) $j_2=20$.
}
\label{fig1}
\end{figure}

All state-to-state cross sections and rate coefficients for quenching 
are available on the UGA Molecular Opacity Project website 
(www.physast.uga.edu/ugamop/). The rate coefficients are also available in
the format of the Leiden Atomic and Molecular Database \citep[LAMDA,][]{sch05}
on our website and the LAMDA site (www.strw.leidenuniv.nl/$\sim$moldata/).
Fig.~\ref{fig1} shows typical examples of state-to-state cross sections 
for $j_2=5$ and 20 for collisions with para-H$_2$. A number of resonances are
evident between 10$^{-2}$ and 10$^{2}$ cm$^{-1}$ which result in significant
modulation in the rate coefficients. However, the magnitude of the
resonances are seen to decrease with $j_2$. Cross sections for some other
$j_2$ can be found in \citet{yang061,yang062}. The agreement 
between the CC and CS calculations is shown to be excellent with a difference
typically better than $\sim$20\%, justifying the adoption of the CS approximation at the higher
energies. 

\begin{figure}
\epsscale{0.9}
\plotone{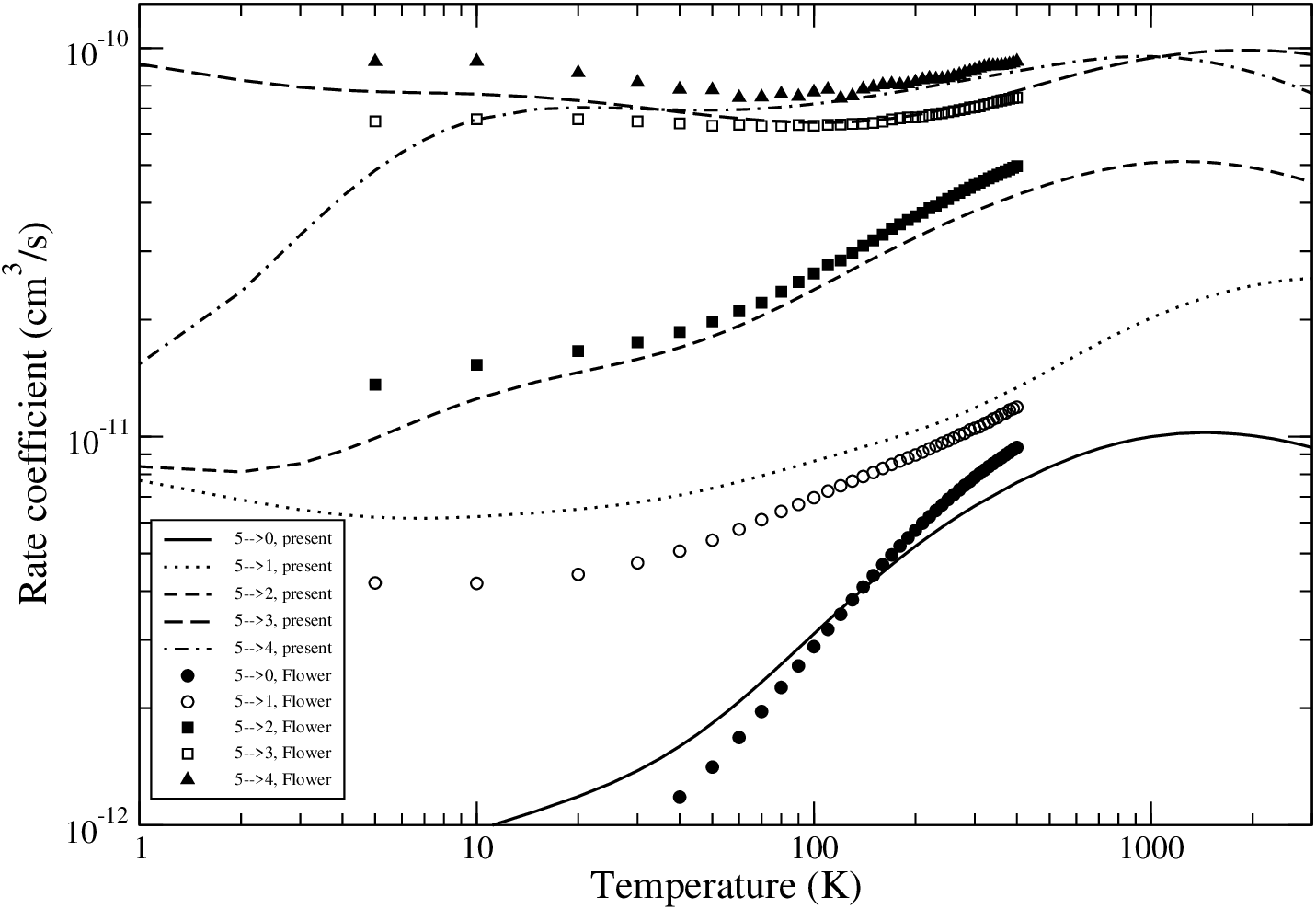}
\caption{
    Rate coefficients for the quenching  of CO($j_2$=5)
    by collisions with para-H$_2$ as
    functions of temperature. Lines indicate present calculations on
    potential V$_{04}$, symbols denote the results of \citet{flower01}
    on  potential V$_{98}$.
    Solid line and solid circles:           $j_2=5\rightarrow j_2^{\prime}=0$,
    dotted line and open circles:          $j_2=5\rightarrow j_2^{\prime}=1$,
    dashed line and solid squares:          $j_2=5\rightarrow j_2^{\prime}=2$,
    dot-dash line and solid squares:      $j_2=5\rightarrow j_2^{\prime}=3$,
    double-dot-dash line and solid triangles:  $j_2=5\rightarrow j_2^{\prime}=4$.
}
\label{fig2}
\end{figure}

\begin{figure}
\epsscale{0.9}
\plotone{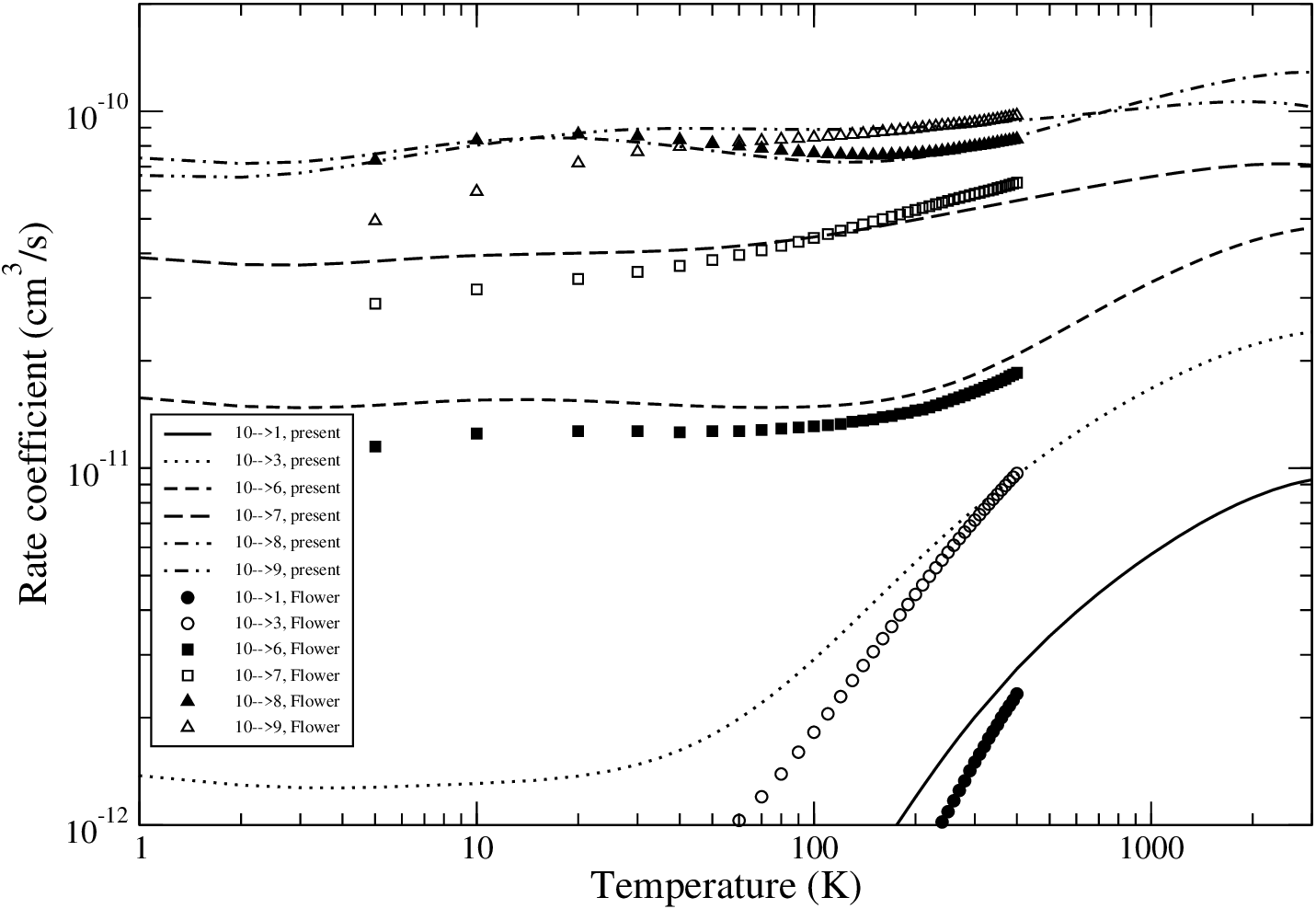}
\caption{Same as Fig.~\ref{fig2}, except for $j_2=10$ and $j_2^\prime$ as indicated.
}
\label{fig3}
\end{figure}

\begin{figure}
\epsscale{0.9}
\plotone{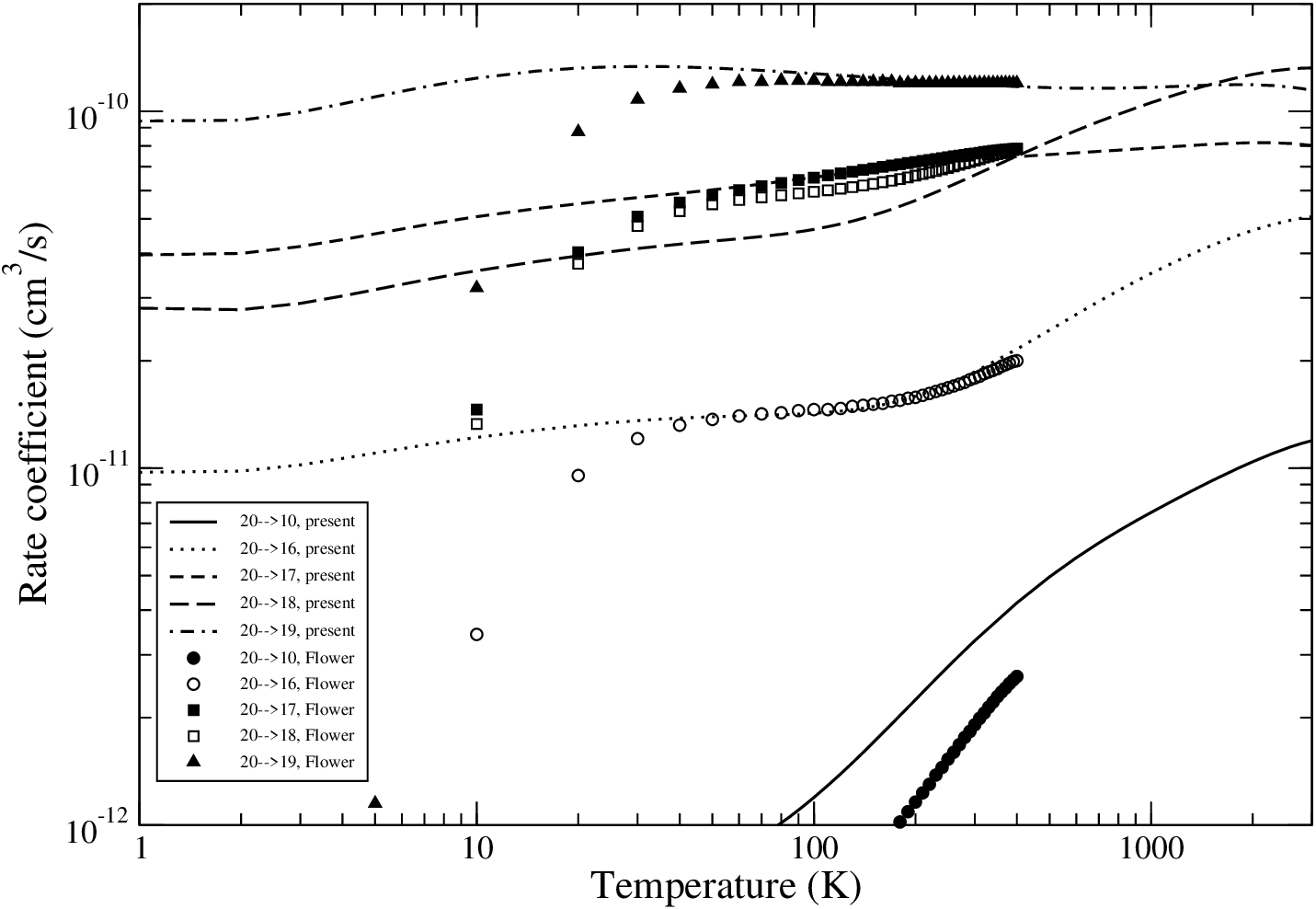}
\caption{Same as Fig. \ref{fig2}, except for $j_2=20$ and $j_2^\prime$ as indicated.
}
\label{fig4}
\end{figure}

\begin{figure}
\epsscale{0.9}
\plotone{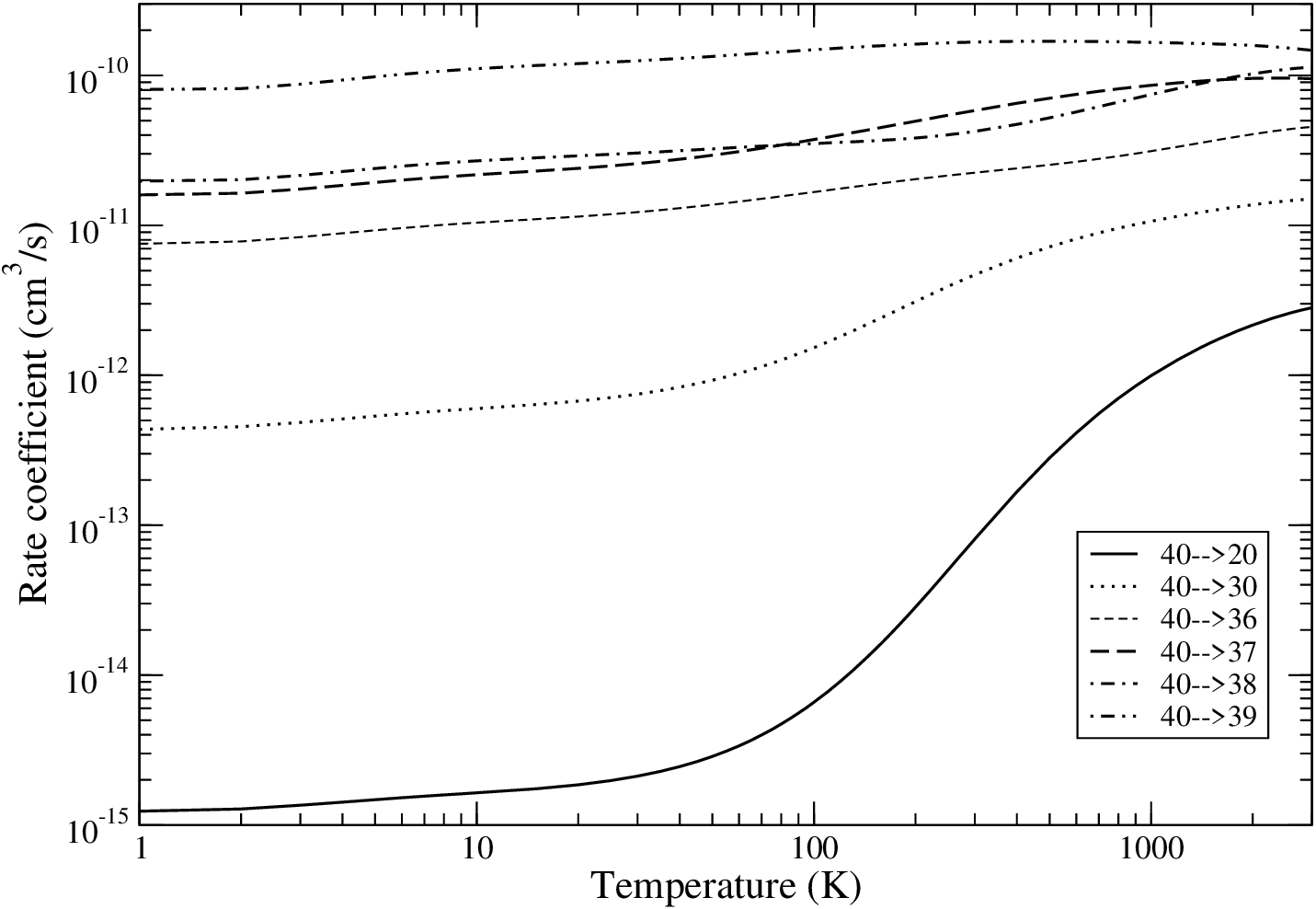}
\caption{Same as Fig. \ref{fig2}, except for $j_2=40$ and $j_2^\prime$ as indicated.
}
\label{fig5}
\end{figure}

\begin{figure}
\epsscale{0.9}
\plotone{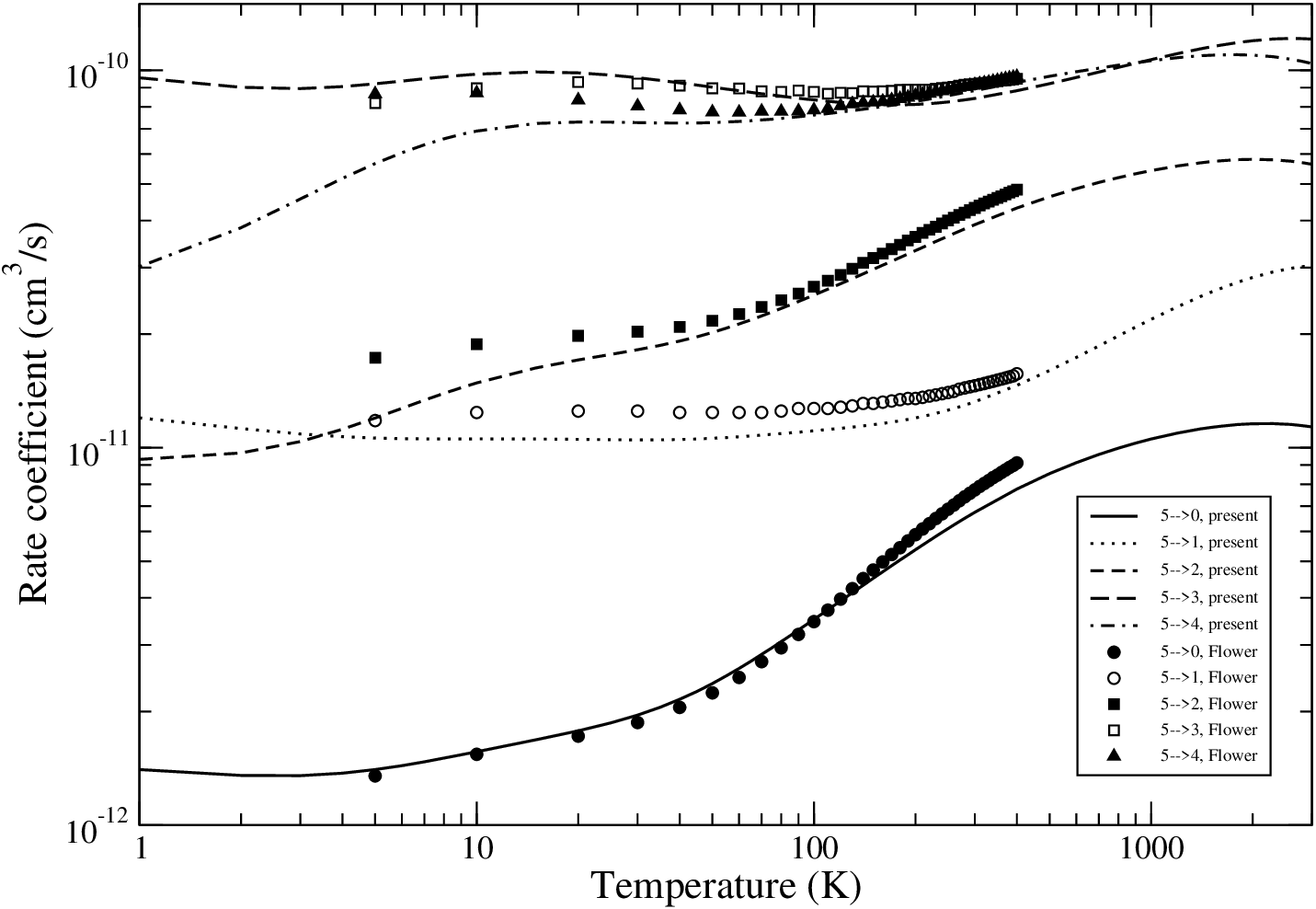}
\caption{Same as Fig. \ref{fig2}, except for ortho-H$_2$.}
\label{fig6}
\end{figure}

\begin{figure}
\epsscale{0.9}
\plotone{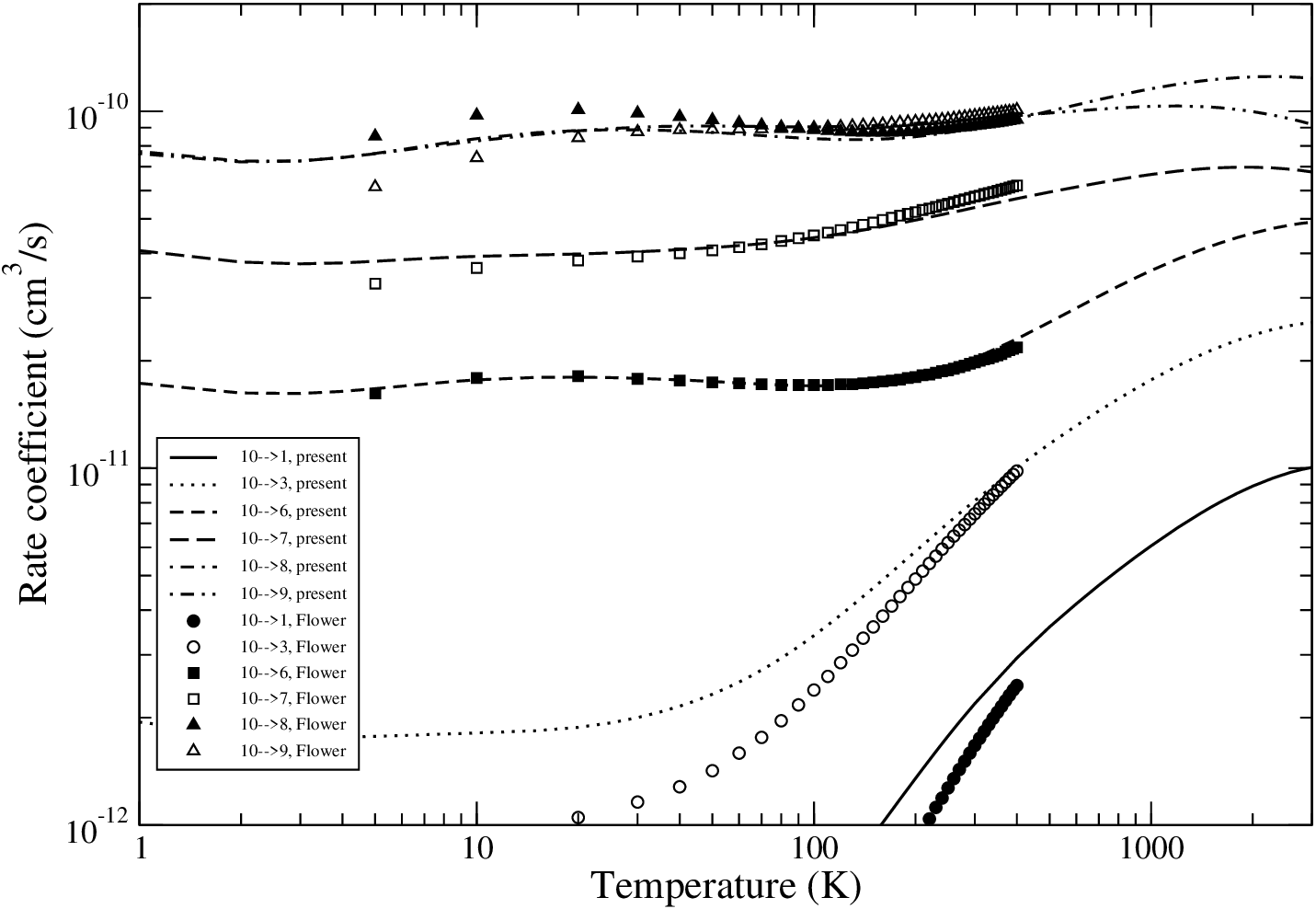}
\caption{Same as Fig. \ref{fig3}, except for ortho-H$_2$.}
\label{fig7}
\end{figure}

\begin{figure}
\epsscale{0.9}
\plotone{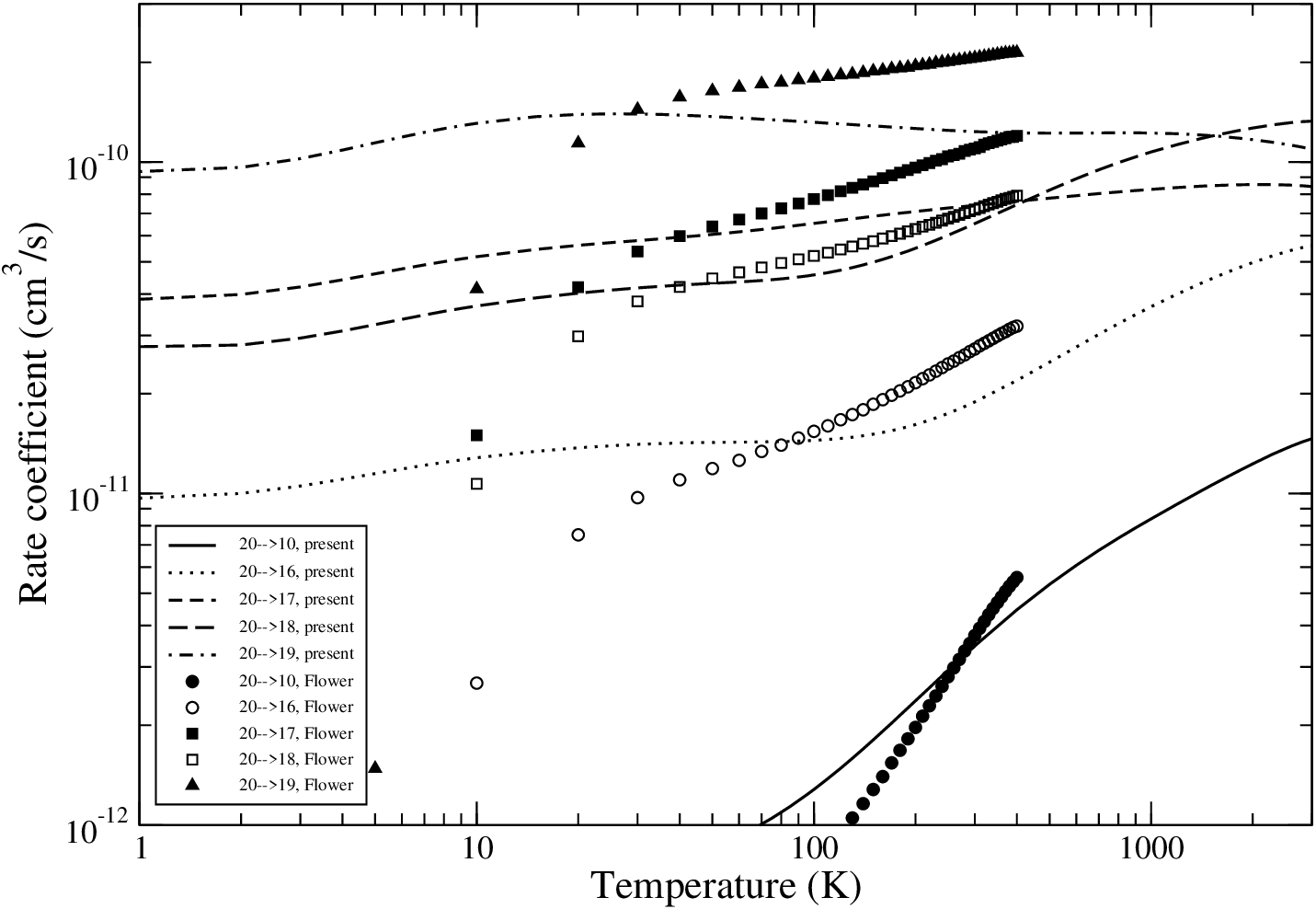}
\caption{Same as Fig. \ref{fig4}, except for ortho-H$_2$.}
\label{fig8}
\end{figure}

\begin{figure}
\epsscale{0.9}
\plotone{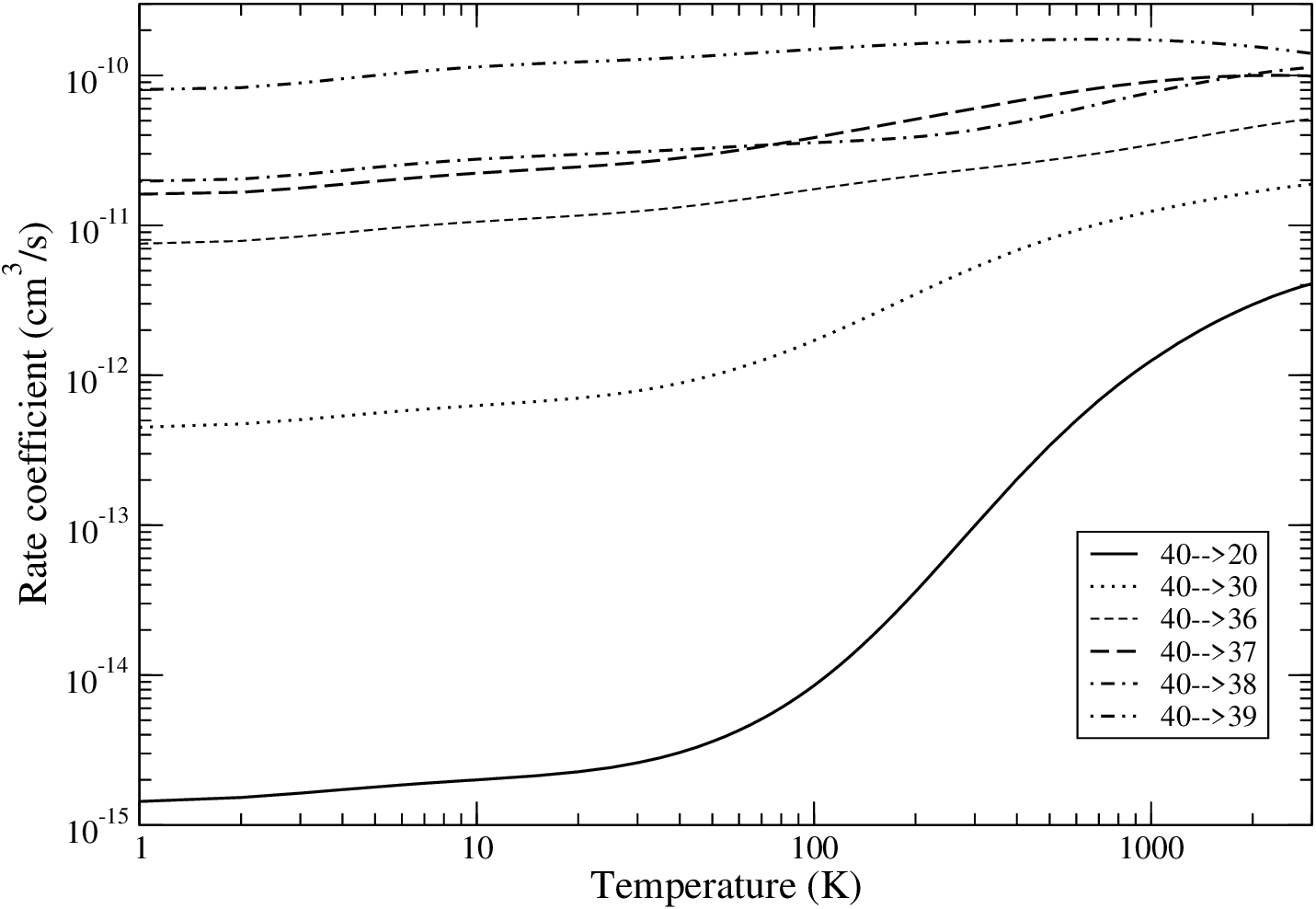}
\caption{Same as Fig. \ref{fig5}, except for ortho-H$_2$.}
\label{fig9}
\end{figure}

Quenching rate coefficients from initial rotational states,
$j_2$=5, 10, 20, and 40, at temperatures ranging from 1 
to 3000 K are shown in Figs. \ref{fig2}$-$\ref{fig9} for CO scattering with para- and
ortho-H$_2$.  Unfortunately, we are unaware of  any
experimental rate coefficient data for these initial states. Therefore,  the current 
results are compared  with the theoretical rate coefficients of \citet{flower01}, 
which were obtained over the limited temperature range, 5 to 400~K, with the 
V$_{98}$ PES. Similar results with V$_{04}$ were obtained by 
\citet{wern06} for rotational levels of CO up to 5 and temperatures in the
range 5$-$70 K.  Wernli {\it et al.} also  presented an analytic fit valid in the
same temperature range. The rate coefficients calculated using their analytic relation are
almost identical to the current results. 

For the quenching of  $j_2$=5 due to para-H$_2$ collisions, 
 Fig.~\ref{fig2} shows that for temperatures between $\sim$1 and 100~K,
which is the van der Waals interaction-dominated regime, the
rate coefficients exhibit an undulatory temperature dependence for 
$\Delta j_2=-1$, $-2$, and $-3$ transitions due to the presence of 
resonances,\footnote{Additional typical examples of scattering resonances in cross section 
can be found in \citet{yang061,yang062} along with
rate coefficient comparisons for $j_2 < 4$.} though the magnitude of the
undulations decrease with increasing $j_2$.
At temperatures above $\sim$100~K, the rate coefficients generally
increase with increasing temperature.
Comparison with the rate coefficients of Flower
shows that at temperatures higher than $\sim$50~K, there is
generally good agreement, except for $j_2'=0$ and 2, where Flower's results 
get larger at temperatures above about 200 K.
The results of Flower are smaller than the present rate coefficients for lower
temperatures with the discrepancy increasing with decreasing temperature.
Exceptions to this behavior are the transitions $j_2=5\rightarrow j_2'=2$ and 4,  
in which Flower's rate coefficients are generally larger than the  present results.
For quenching from initial levels $j_2=$10 and 20 given in Figs.~\ref{fig3} and \ref{fig4}, respectively,
the state-to-state rate coefficients
have similar structure to that shown for $j_2=5$. The rate coefficients are dominated by small $\Delta j_2$. 
For $\Delta j_2=-1$ transitions, a slight bump can be seen at temperatures between 
$\sim$1 and 100~K for $j_2=10$ and 20 as shown in Figs.~\ref{fig3} and \ref{fig4}. 
Figure~\ref{fig5} displays
quenching rate coefficients for $j_2=40$. No previous results are available as the
calculations of \citet{flower01} stopped at $j_2=29$ for para-H$_2$ collisions.

For scattering by ortho-H$_2$, the trends noted for para-H$_2$ are also evident
as displayed in Figs.~\ref{fig6}-\ref{fig9}.
The quenching rate coefficients are of similar magnitude as those
obtained for para-H$_2$ and show very similar structure. The highest level 
considered by \citet{flower01} for ortho-H$_2$ collisions was $j_2=20$.

Comparing Figs.~\ref{fig2}-\ref{fig9}, some general trends can be noted. With few exceptions,
the state-to-state rate coefficients appear to attain 
constant values as they decrease to 1~K.
However, this is deceptive as an orbiting resonance is typically present in
the cross sections for collision energies between $\sim$0.01-1 cm$^{-1}$ (see Fig.~\ref{fig1})
computed
on the V$_{04}$ PES \citep{yang061}. For temperatures less than shown,
the rate coefficients typically decrease (increase) prior to the Wigner
regime \citep{Wigner48} for intermediate (high) $j_2$ before eventually attaining 
finite limiting values, typically below $\sim$1 mK.

Disregarding the effect of the resonances, one would expect two possible trends
in the relative  ordering of the rate coefficient values as a consequence of
energy and angular momentum gaps. The exponential energy gap
law \citep{ste91} predicts that the rate coefficients should decrease with increasing internal energy
difference between the initial and final levels (i.e., with increasing $|\Delta j_2|$).
However, CO is nearly homonuclear and rotational transitions in homonuclear
molecules follow an even $\Delta j_2$ propensity rule \citep[e.g.,][]{she07},
though the effect may diminish for large values of $|\Delta j_2|$. Both effects appear
to be important for CO-H$_2$. For example, $\Delta j_2=-2$ is the dominant
transition for quenching for  temperatures greater than $\sim$1000-3000~K. It is
also the dominant transition for temperatures less than $\sim$10~K, but only
for $j_2=5$ and 10.  For other temperatures, $\Delta j_2=-1$ transitions are predominant, 
while for some cases $\Delta j_2=-3$ is secondary, but with a rate coefficient
larger than the $\Delta j_2=-2$ transitions. However, for $\Delta j_2<-3$, the
exponential energy gap law appears to be fulfilled in nearly all considered
cases.

\section{ASTROPHYSICAL IMPLICATIONS}
 
As discussed in the introduction, carbon monoxide plays an
important role in a variety of astrophysical and atmospheric
environments and has been detected in countless sources
under a variety of excitation conditions. We discuss a few
of these below focusing primarily on observed and predicted 
transitions from intermediate- and high-$j_2$ in the FIR.

Rotational emission lines of CO have been observed in high-redshift ($z$)
objects including the quasar SDSS J1148+5251 at $z=6.42$ using
the Very Large Array. \citet{nar08} modeled
the emission from CO($v=0,j_2$) with $j_2=1-10$ with a non-LTE radiative
transfer code and considered excitation due to H$_2$ with rate
coefficients taken from the LAMDA database. They found that the
CO spectral energy distribution (SED) peaks at $j_2=8$ at the beginning
of the quasar phase, $z\sim 7-8$. Observations and modeling of the
CO emission can serve to probe the hierarchical buildup of the
host galaxy and the quasar phase and can be used to estimate the
halo mass and quasar host morphology.

\begin{figure}
\epsscale{0.9}
\plotone{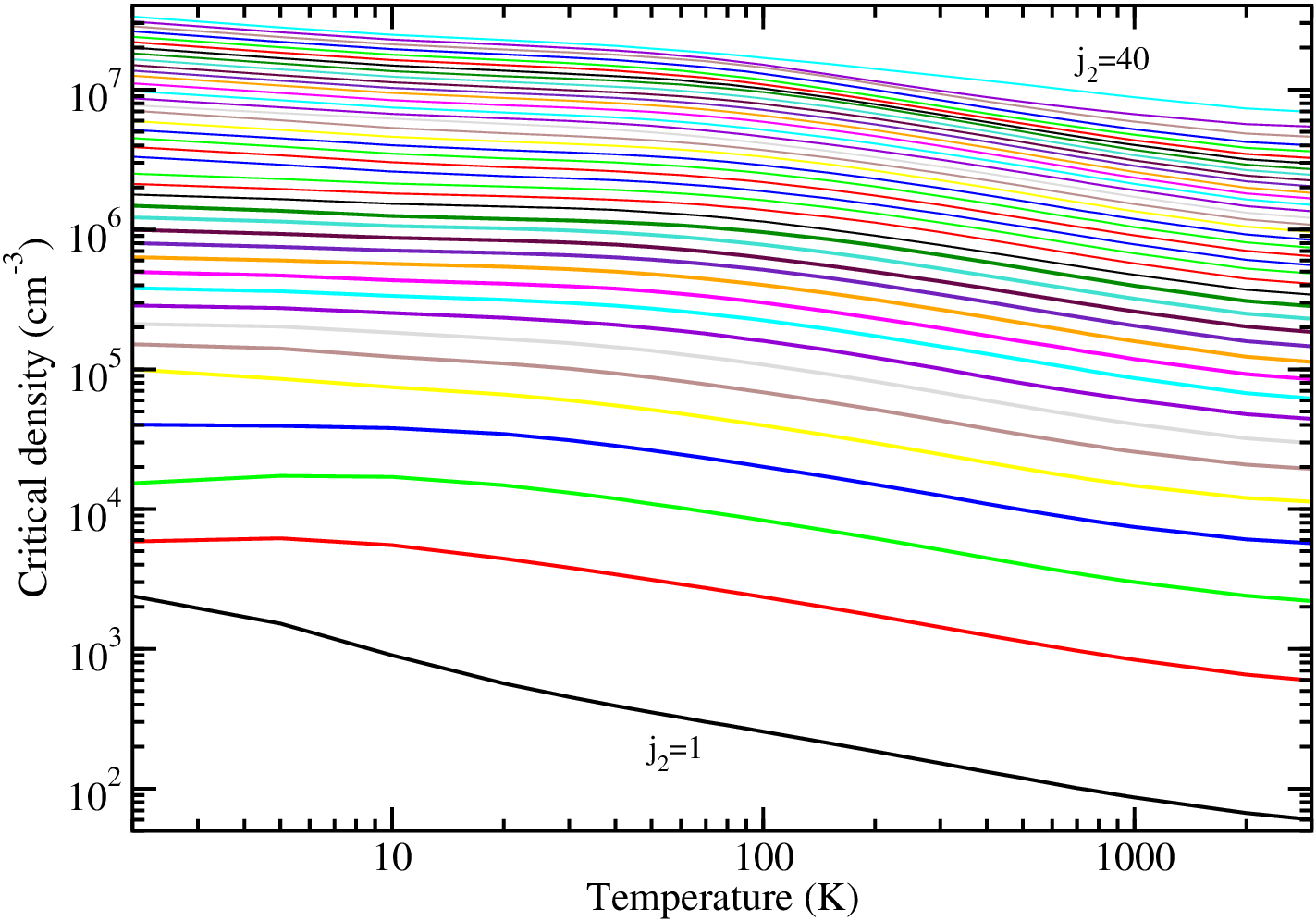}
\caption{Critical densities for CO($v=0,j_2$) due to para-H$_2$
collisions as a function of gas temperature $T$.}
\label{fig10}
\end{figure}

\citet{spa08} proposed that
 CO emission from $j_2$ as high as 30  could be
observed from high-$z$ massive accreting black holes using
the Atacama Large Millimeter/submillimeter Array (ALMA). 
Such observations could provide a diagnostic of
the radiation source: PDR or x-ray dominated region (XDR), for example. However, as the
densities of such objects are typically 10$^5$ cm$^{-3}$, collisions
with the dominant species, H$_2$, will determine the CO rotational
populations. In Fig.~\ref{fig10}, the critical densities for CO due to
para-H$_2$ collisions are plotted (neglecting optical depth effects) where the
critical density of rotational level $j_2$ is defined as
\begin{equation}
n_{\rm cr}(j_2) = {{\sum_{j_2'<j_2} A_{j_2\rightarrow j_2'}}\over{\sum_{j_2'\neq j_2}
k_{j_2\rightarrow j_2'}(T)}}
\end{equation}
\citep[e.g.,][]{ost06} and $A_{j_2\rightarrow j_2'}$ are the spontaneous transition probabilities for
dipole transitions.
Rotational levels $j_2\gtrsim 10$ are
seen to be clearly out of thermal equilibrium for a gas density of
10$^5$ cm$^{-3}$ requiring a non-LTE
analysis. As the LAMDA database incorporates the CO-H$_2$ 
rate coefficients of \citet{wern06} and \citet{flower01} with extrapolations
for para- and ortho-H$_2$ collisions above $j_2=29$ and 20, respectively,
the predicted CO line intensities of \citet{spa08} for emission from
levels larger than $j_2=20$ would likely be improved if the current
rate coefficients are adopted.
Interestingly, their predicted CO SEDs  peak at
$j_2=10$ and $j_2=25$ for PDR and XDR environments, respectively.

The $j_2=6\rightarrow 5$ and $j_2=7\rightarrow 6$ lines have been observed in the
starburst galaxy NGC 253 \citep{hai08,bra03}. Using a non-LTE analysis, \citet{bra03}
concluded that the cooling in the CO lines is so large that it must be balanced
by a cosmic ray heating rate $\sim$800 times greater than that in the Milky Way.
Previously, \citet{kro89} suggested that highly-excited $j_2$ lines (e.g., $j_2=16-58$)
could be detectable from Seyfert galaxies giving diagnostics of the internal pressure of
the dust-obscured torus.

The {\it Infrared Space Observatory} studied a large variety of pre-main sequence
objects (i.e., protostars) and detected numerous FIR CO rotational lines.
For example, emission lines from $j_2=14-26$ were observed in the Class 0 object
L1448-mm \citep{nis99}, $j_2=14-31$ in the Class I object IC 1396N, $j_2=14-16$
in the Class I object W28 A2, $j_2=14-20$ in the Class II object R CrA \citep{sar99},
and $j_2=14-25$ in T Tauri \citep{spi99}.  Non-LTE analyses of these pre-main
sequence objects suggested multiple components with various excitation
conditions. Future studies of these protostars, the infrared sources discussed above,
and future FIR and submillimeter observations with {\it Herschel} and
ALMA can benefit from the CO-H$_2$ quenching rate coefficients presented
here.

\section{CONCLUSIONS}

Rotational quenching of CO due to para- and ortho-H$_2$ collisions has 
been studied using an explicit  quantum-mechanical close-coupling approach
and the coupled-states approximation on the potential surface, V$_{04}$, 
of \citet{jankow05}.  State-to-state quenching rate coefficients 
for initial rotational levels $j_2=$1, 2, $\cdots$, 40 of CO
are obtained over a wide temperature range and available in tables formatted
for astrophysical applications.  Resonances result in 
undulations in the temperature 
dependence of the rate coefficients with the amplitudes of the
undulations typically decreasing with $j_2$. For temperatures less than $\sim$50~K,
the current state-to-state rotational quenching rate coefficients
obtained with V$_{04}$ are found to depart from the results of \citet{flower01}
obtained with the earlier V$_{98}$ PES.  
The discrepanices are likely related to the differences in the well
depth and anisotropy
of the two potentials. 

\begin{acknowledgements}
BHY and PCS acknowledge support from  NASA grant NNX07AP12G.
NB acknowledges support from NSF through grant PHY-0855470.
RCF acknowledges
support from NSF through grant PHY-0854838.
We thank Dr. Piotr Jankowski for providing his CO-H$_2$ potential subroutine
and Prof. Gary Ferland and Dr. Donghui Quan for discussions on astrophysical applications.
Part of calculations were performed on EITS/RCC pcluster at the University of Georgia.
\end{acknowledgements}

\end{document}